\documentclass[preprints,review,accept,moreauthors]{Definitions/mdpi}
\firstpage{1}
\pubvolume{xx}
\issuenum{1}
\articlenumber{5}
\pubyear{2020}
\copyrightyear{2020}
\history{} 

\Title{Exclusion zone phenomena in water - a critical review of experimental findings and theories}

\Author{Daniel C. Elton$^{1 *}$\orcidA{}, Peter D. Spencer$^{2 *}$\orcidB{}, James D. Riches$^{3}$, Elizabeth D. Williams$^{4}$\orcidC{}}

\AuthorNames{Daniel C. Elton,Peter D. Spencer,James D. Riches,Elizabeth D. Williams}

\address{%
$^{1}$ \quad Radiology and Imaging Sciences, National Institutes of Health Clinical Center, Bethesda, MD 20892, USA\\
$^{2}$ \quad School of Biomedical Sciences, Faculty of Health, Institute of Health and Biomedical Innovation, Queensland University of Technology (QUT), Australia\\
$^{3}$ \quad School of Earth, Environmental and Biological Sciences, Science and Engineering Faculty, Institute for Future Environments, QUT, Australia\\
$^{4}$ \quad School of Biomedical Sciences, Faculty of Health, Institute of Health and Biomedical Innovation, QUT, Australian Prostate Cancer Research Centre - Queensland, Translational Research Institute, Australia
}

\corres{Correspondence: daniel.elton@nih.gov, p.spencer@hdr.qut.edu.au}

\abstract{The existence of the exclusion zone (EZ), a layer of water in which plastic microspheres are repelled from hydrophilic surfaces, has now been independently demonstrated by several groups. A better understanding of the mechanisms which generate EZs would help with understanding the possible importance of EZs in biology and in engineering applications such as filtration and microfluidics. Here we review the experimental evidence for EZ phenomena in water and the major theories that have been proposed. We review experimental results from birefringence, neutron radiography, nuclear magnetic resonance, and other studies. Pollack and others have theorized that water in the EZ exists has a different structure than bulk water, and that this accounts for the EZ. We present several alternative explanations for EZs and argue that Schurr's theory based on diffusiophoresis presents a compelling alternative explanation for the core EZ phenomenon. Among other things, Schurr's theory makes predictions about the growth of the EZ with time which have been confirmed by Florea et al.\ and others. We also touch on several possible confounding factors that make experimentation on EZs difficult, such as charged surface groups, dissolved solutes, and adsorbed nanobubbles.}


\begin{document}



\section{Introduction}
Prof. Gerald Pollack's group has provided many convincing experimental demonstrations of an exclusion zone (EZ) in water whereby particles such as plastic microspheres are repelled from a surface. The width of the EZ depends on the properties of the surface and ambient conditions and may reach hundreds of microns. In addition to small particles, there is evidence that the EZ excludes relatively large molecules such as pH-indicators and biological molecules. 

For the case of highly hydrophillic surfaces these findings have now been reproduced by several independent research groups\cite{Chen2011,Musa2013,Huszr2014,Gudkov2014,Jabs2014, Florea2014,bunkin2013,Yakhno2018,Spencer2018,Sharma2018,Esplandiu2020} and constitute a genuine physical phenomena which is in need of a theoretical explanation. A few experimenters have reported EZs near metal surfaces,\cite{Zheng2006EZ,Pedroza2015,Chai2012metal} and one has reported it for cellulose.\cite{Sulbarn2014} In this work we present a review of exclusion zone phenomena, including many recent experimental studies, and describe several mechanisms by which the EZ phenomena can arise. In any given experimental scenario, some or all of those mechanism may be present. EZ phenomena may have important engineering applications in water filtration, reducing biofouling,\cite{PedregalCorts2019} and microfluidics.\cite{Florea2014} EZ phenomena also have obvious importance to understanding biological systems and resolving outstanding questions about ``biological water''.\cite{Ball2008} 



\section{Background}
The existence of structured water near hydrophilic interfaces has been proposed several times previously. Drost-Hansen (1969, 1973) reviewed many experiments and came to the conclusion that interfacial (``vicinal'') water exhibits structural difference that extend to tens to thousands of molecular diameters.\cite{DrostHansen1969, DrostHansen1973} A common theme found in the literature is that hydrophillic surfaces result in a change in the structure of interfacial water which amounts to ``templating'' of the surface.\cite{Fenter2004, Verdaguer2006, Maccarini2007} Many claims for ordering near biological interfaces (ie.\ in cells or small blood vessels) have been made, with many positing that ``biological water'' has significant structural differences.\cite{Ling2003}  One of the earliest studies in this vein was performed by Deryagin in 1986, who also described an EZ type phenomena in cells.\cite{Florea2014, Deragin1986} A difficulty in such research is  separating out property changes that occur due to confinement, which are largely thermodynamic in nature (ie.\ from Laplace pressure), from effects due to the putative restructuring of cellular water. Despite many works on ``biological water'', the hypothesis that cellular water undergoes significant restructuring remains very controversial (for a review, see Ball, 2008).\cite{Ball2008} It is not our intent to review that controversy here, but only to highlight its relationship to EZ water. 

At a hydrophilic surface, the alignment of hydrogen bonds at the surface may create a polarized layer and electric field, the influence of which may extend out for several layers of water molecules. This argument has been used to support both experimental evidence from X-ray and spectroscopic studies for order at the water-hydrophilic surface interface.\cite{Shen2006,Catalano2011,EftekhariBafrooei2010,Fenter2004}. While this ordering is often called ``long-range'', the extend found in most studies is only a few water layers (ie.\ 1-2 nm). This level of restructuring, which extends just a few molecular layers, is consistent with the predictions of double layer theory\cite{Fenter2004} and molecular dynamics studies quantifying the extent of angular correlation in the bulk and near interfaces.\cite{Ebbinghaus2007,Elton2014,EltonThesis} The limited extent of restructuring is not surprising given that hydrogen bonds are relatively weak (0.24 eV per bond) and are short lived due to thermal perturbations (lifetime $\approx$1 ps).\cite{Chaplin2000,EftekhariBafrooei2010}

Moving beyond structural changes, it has been shown that ion exchange membranes such as Nafion (heavily studied by Pollack and discussed below) can introduce electrical changes.\cite{Park2006} These changes have been evidenced by Electrical Impedance Spectroscopy, which measures the electrical potential within a system by passing an alternating current of known frequency and small amplitude through it.\cite{Park2006}  

\section{Pollack's key experimental findings and replications}
\begin{table}[h]
    \caption{Some of the observed properties of the exclusion zone water}
    \begin{tabular}{lccl}
measured property  & EZ water value & bulk value & references     \\
\hline
refractive index      & 1.46 &  1.33 & Bunkin et al., 2013 \cite{bunkin2013}\\
T$_2$ relaxation time & 27.2$\pm$0.4 ms  &  25.4 $\pm$ 1 ms & Zhen et al, 2006\cite{Zheng2006EZ}\\ 
electric potential near surface &  -120 to -200 mV  & 0 mV & \cite{Zheng2006EZ,Zheng2009ordered,Klimov2007}\\ 

     \end{tabular}\label{EZproperties}
\end{table}

The exclusion zone was first described by Pollack et al.\ in 2003 after they observed latex microspheres in suspension moving away from the surface of the hydrophilic material Nafion (a sulfonated tetrafluoroethylene based fluoropolymercopolymer developed by DuPont) under a microscope.\cite{Zheng2003} Using UV-vis absorption spectra and NMR, in 2006 Pollack et al.\ argued that EZ water exists in a different phase.\cite{Zheng2006EZ} Further investigations from Pollack's lab in 2007 using microelectrodes indicated that the EZ region is negatively charged.\cite{Klimov2007} Introduction of pH sensitive dye indicated a low pH ($<$3) close to the Nafion surface, as well as a small region very close to the surface where the dye appeared to be excluded.\cite{Chai2009} On the other hand, experiments by Chai, Mahtani, and Pollack (2012) showed that EZs near the charged surfaces of some metals are positively charged.\cite{Chai20122} Additionally, water in the EZ was reported to have a higher index of refraction, which is attributed to a higher density due to a change in the water's structure.\cite{pollack2013fourth} Hwang et al.\ attempted to measure the increase in density by dissolving a hydrophilic ceramic powder in water and then filtering the water, but only a small (0.4\%) increase was observed.\cite{Hwang2018} While most experiments have been done showing exclusion of microspheres, one experiment from Pollack's lab shows rejection of salt as well.\cite{ZhangWATER2015} A summary of properties that have been reported for EZ water can be found in table \ref{EZproperties}. 
 
\begin{figure*}[h]
\centering
    \includegraphics[width=10cm]{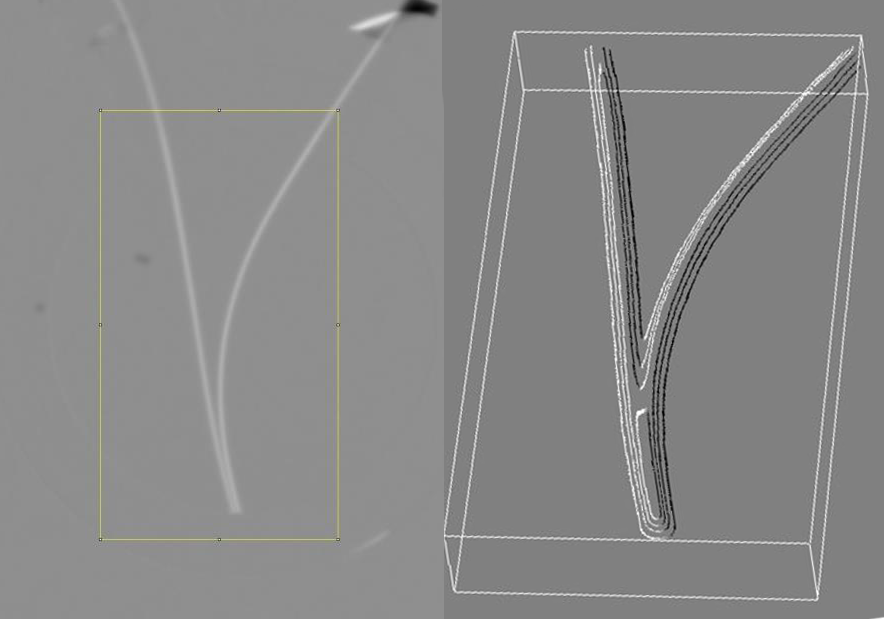}
    \caption{Image produced by subtracting the natural logarithm of the neutron attenuation in the distilled water filled cell with and without two strips of Nafion. The yellow outline shows the region of interest for creating the 3D surface plot shown on the right. }
    \label{neutronatten}
\end{figure*}   

\section{The structure change theory}
A few researchers have proposed that the EZ is due to a change in water's structure.\cite{Zheng2006EZ,Zheng2009ordered,Figueroa2011,Oehr2014,Giudice2013,Elia2019,Giudice2010} One problem with this is there are no obvious thermodynamic forces in the system to drive such a phase change or long range ordering. In his popular science book ``The Fourth Phase of Water'', Pollack hypothesizes that the EZ water is structured in hexagonal sheets, with the hydrogens lying directly between oxygens.\cite{pollack2013fourth}  Pollack proposes that when these sheets are stacked hydrogen atoms bond to the oxygens in neighboring layers, such that each hydrogen forms three bonds. It is important to note that his book is not peer reviewed and not a scientific monograph, and Pollack admits that the idea of a layered structure is speculative.\cite{pollack2013fourth} In other work, Pollack has proposed that the structure is an intermediate between ice and water.\cite{So2011} Oehr and LeMay (2014) propose that EZ water may comprise tetrahedral oxy-subhydride structures.\cite{Oehr2014} 
It is worth noting that in 1962 Fedayakin proposed that ``polywater'' (discussed below) had a similar honeycomb like structure with each oxygen bonded to 3 hydrogens.\cite{Fedyakin1962} In 1971 Hasted noted problems with hexagonal water structures in general, noting that high energy cost of placing hydrogens between oxygens was enough to make such a structure explode if it were ever created.\cite{Hasted1971} Much more recently  Seggara-Mart\'{i} et al.\ performed quantum chemistry calculations showing such a structure to be unstable.\cite{SegarraMart2014} Further quantum chemistry calculations were performed on two stacked hexagonal layers (each layer contained two hexagons and one negative change (H$_{19}$O$_{10}^{-}$). The negative charge did not distribute uniformly over the structure and optimization of the structure resulted in a ``bulk-type water aggregate'', showing it to be unstable.\cite{SegarraMart2014MDPI} 

Elia et al. suggest that perturbations near the EZ surface can cause clumps of EZ water to disperse in the bulk liquid, resulting in changes that can be detected in the bulk liquid as the EZ water clumps dissipate.\cite{Elia2019} If such clumps do indeed exist their quasi-stability would provide evidence for the structure change theory. Figuera and Pollack have presented a somewhat similar argument, arguing that the stable nature of the EZs under perturbations must be due to a structure change in the liquid.\cite{Figueroa2011} Guiduce et al. propose a phase transition occurs in the EZ which can be understood using quantum electrodynamic calculations.\cite{Giudice2010,Giudice2013,Giudicebook}

Exclusion zone phenomena have been observed in other polar liquids as well such as dimethyl sulfoxide (DMSO), suggesting that hydrogen bonds are not required for the phenomena.\cite{Chai2010} If it were the case that EZs were due to a phase change we would expect EZ phenomena would be quite different between water, which supports low density hexagonal structures and hydrogen bonding, and other polar solvents which do not. As we discuss in the next section, neutron radiography does not support the notion of a higher density phase. An experiment which could shed additional light on this subject is X-ray crystallography. X-ray crystallography has not been done for the EZ but has been used to examine the electrically-induced water bridge which Pollack hypothesizes may be made of EZ water in his popular science book \textit{The Fourth Phase of Water: Beyond Solid, Liquid, and Vapor} (2013). Both molecular dynamics simulation,\cite{Skinner2012} X-ray crystallography,\cite{Skinner2012} and neutron scattering\cite{Fuchs2009neutron} show that the internal structure of the water bridge is unchanged - implying that it is supported by enhanced surface tension rather than a change in internal structure. 

Pollack points to enhanced absorption at 270 nm as evidence for a possible phase change in the EZ.\cite{Zheng2006EZ,So2011} This absorption peak was not found in quantum chemistry simulations.\cite{SegarraMart2014MDPI} Strikingly, results from Pollack's own lab show that a similar absorption peak is seen in pure salt solutions (LiCl, NaCl, KCl), so the source of this enhanced absorption appears to be related to dissolved solutes.\cite{Chai2008absorption} A study of Arrowhead Spring found absorption at 270 nm in bulk water.\cite{DibleWATER2014} Hypothesizing that EZ water would be a transitionary form between ice and liquid water, Pollack performed IR measurements of melting ice.\cite{So2011} During the course of these experiments the 270 nm peak sometimes (but not always) appeared transiently (ie.\ for a few seconds) while the ice was melting. In the same work they also report that degassing the water (either through boiling, drawing a vacuum, or nitrogen bubbling) reduced of the appearance of the peak.\cite{So2011} Thus, it's also possible that the peak is related to tiny bubbles trapped in the ice which migrate to the surface while the ice is melting. A possible mechanism which may be involved here is that the absorption is from superoxide anions (O$^{-2}$) and their protonated form, the hydroperoxyl radical (HO$_2$). These two species exist in equilibrium in small quantities when oxygen is dissolved in water, exist in much higher quantifies in acidic solutions (such as near Nafion, see below), and can be induced by UV radiation.\cite{Bielski1985} There is evidence that both species absorb in the range of 240-260 nm.\cite{Janik2013,Bielski1985,Matheson1969,Czapski1964}

Pollack also hypothesizes that when light is shined on EZ water it causes positive and negative charges to separate, and the EZ water region to grow.\cite{Chai2009} This is problematic since water is a good conductor and charge separation would be difficult to sustain. In his book Pollack proposes that blood flow is powered by EZ phenomena.\cite{pollack2013fourth} This idea is in contradiction with the fact that mammals can live in darkness (for instance the naked mole rat) and often have heavy black fur which blocks out most sunlight. Conveniently, Pollack's proposed layered structure dovetails nicely into a long history of companies selling ``structured'' or ``hexagonal'' water for health purposes. Tests of some of these products with nuclear magnetic resonance spectroscopy (NMR) show no difference from pure water.\cite{shinNMRweb} Companies currently selling EZ water products for health include include Divinia Water, Structured Water Unit LLC, Flaska, Advanced Health Technologies (vibrancywater.ca), and Adya Inc. The idea of utilizing EZ water for health has been promoted by influential figures in alternative medicine such as Dr.\ Joseph Mercola and Dave Asprey. Scientific skepticism is called for here as the potentially relevant experiments from Pollack's lab have not been replicated and no studies have shown any benefit of EZ water to health in animals or humans. Instead of providing words of caution, however, Pollack has embraced the attention he has received from alternative medicine community by participating in podcasts with Mercola, Asprey, and many others where he promotes the idea that EZ water is important for health. While Pollack has been careful to leave his ideas about EZ water and human health out of his publications, several studies which explore his ideas have appeared in peer reviewed journals, thus making this topic relevant for this review.\cite{Pokorn2015:675,Sharma2018,Hwang2017,Tychinsky2014}  

\subsection{Testing the structure change theory with neutron radiography}
\label{sec:neutron}

As described in detail in \cite{Spencermastersthesis}, some of the authors on this work recently undertook a neutron radiography study to measure the density of water near the Nafion surface. Pollack's proposed EZ water structure has a density which is $\approx$ 10\% higher than liquid water.  Neutron radiography has previously been used to measure subtle density differences between supercritical and subcritical water.\cite{Takenaka2013} The experiment was conducted using the Dingo radiography imaging station at the Australian Nuclear Science and Technology Organization (ANSTO). The neutron flux varied between 1.14 x 10$^7$ to 4.75 x 10$^7$ neutrons cm$^2$ s$^{-1}$. Imaging with test objects indicated the instrumental resolution was at least 100 $\mu$m, which is adequate to detect an EZ extent of 200 $\mu$m, smaller than the extent of 500+ $\mu$m proposed by Pollack and collaborators.\cite{Zheng2006EZ,Das2013} In the experiment, a 2 mm wide quartz glass cell was filled with distilled water and two strips of Nafion were inserted. The temperature was held at 21$^{\circ}$ $\pm$ 1 $^{\circ}$C. and the Nafion strips were 0.43 mm thick and 1-2 mm in width. It was expected that a denser region of EZ water would nucleate from the Nafion surface, resulting in greater neutron attenuation. The arrangement of the two Nafion strips in a “V” formation was intended to create an effect where the visible difference due to EZ formation could be doubled, creating an EZ region large enough to be identified between the strips. Figure \ref{neutronatten} shows the difference between the natural logarithm of attenuation in the cell with and without two strips of Nafion. As can be clearly seen, no density differences are observable near the surface, at least within the 100 $\mu$m resolution of the instrument. An NMR study has showed water polarization and ordering next to fused silica (an allotrope of quartz), but the extent of this ordering was found to be limited to 60 molecular layers.\cite{Totland2016} Thus it can be concluded that EZs do not form near quartz to begin with.\cite{Totland2016}

\subsection{Testing for structure change with optical birefringence measurement}
Another piece of experimental evidence that Pollack presents for EZ water having a different structure is the presence of include optical birefringence in the EZ caused by Nafion.\cite{pollack2013fourth,Yoobihan2016water} Attempts to replicate this result was performed by some of the authors using a polarized light microscope setup.\cite{Spencer2018,Spencermastersthesis} In a similar vein, Bunkin et al.\ and Tychinsky have reported an increase in the refractive index of water very close to the surface of Nafion.\cite{bunkin2013,Tychinsky2011} It was found that there are confounding factors which cause the appearance of birefringence near the surface of Nafion. Both air-dried Nafion and zinc still exhibited a high degree of birefringence near the surface due to light reflected obliquely from the surface.\cite{Spencer2018} The way that the surface was cut also changed the degree of reflection birefringence observed, with a blade cut surface showing more of this effect than a rough surface cut with scissors. 
In addition, in some cases microspheres reflect light and thus give the appearance of a wide birefringent region extending from the material surface into the bulk water.\cite{Spencermastersthesis} In a similar vein, polarization by reflection has been noted to play a confounding role in the measurement of the birefringence properties of ice.\cite{icelab2015} Thus, the  measurements of birefringence near the surfaces of Nafion, zinc, and other metals were due to optical effects from uncontrolled-for reflections and do not constitute an evidence for underlying crystalline ordering in water.

\subsection{Aside: historical parallels of the structure change theory with polywater}
Polymeric water (``polywater'') was purported to be a special phase of water which formed when water was condensed into tiny capillary tubes with diameters smaller than 100 micrometers. Interestingly, the structure which was proposed for polywater is very similar to the the structure Pollack proposes for EZ water. The earliest papers on polywater phenomena originated from the group of Boris Deryagin at the Institute of Surface Chemistry in Moscow, USSR in the early 1960s.\cite{Deragin1967} In 1962 Fedayakin proposed that polywater had a honeycomb like structure with each oxygen bonded to 3 hydrogens.\cite{Fedyakin1962} Lectures by Deryagin in England and the United States in 1966, 1967 and 1968 drew the attention of Western researchers. Research interest peaked after a 1969 a paper by Lippincott et al.\ in {\it Science} which reported spectroscopic results which were said to provide conclusive evidence of a ``stable polymeric structure".\cite{Lippincott1969AAAS} Over 160 papers on polywater were published in 1970 alone.\cite{EISENBERG1981Science} However, by 1972 it became apparent that the observed phenomena were due to trace amounts of impurities,\cite{Rousseau1970Science} some of which likely came from human sweat.\cite{Rousseau1971Science} In some cases it was found that the sample tubes contained very little water at all. Altogether, over 500 publications were authored on polywater between 1963-1974.\cite{EISENBERG1981Science, Bennion1976} Far from being just a historical curiosity, the polywater saga is something that EZ water researchers can learn from to avoid repeating the mistakes of the past. The polywater saga is an example of what Langmuir called ``pathological science'', whereby a community fixates on a particular theory while disregarding other explanations. Other features of pathological science are that the experimental evidence is often on the edge of significance, and that interest in the pathological theories persists for years after disconfirming evidence and better theories have been presented. There is a long history of pathological science regarding water, which is probably related to the fact that water's properties can change dramatically under the influence of trace solutes and dissolved gases which are hard to control experimentally. To give another example, the Mpbema effect, where hot water is observed to freeze faster than cold, is now recognized as another case of pathological water science. Invariably the experiments that found such an effect were later shown to potentially plagued by container variation, impurities, dissolved gases, and unwanted evaporation. The most carefully controlled experiments (Brownridge, 2011) have shown the only differences are due to unavoidable variations in the nucleation sites in identical glass containers.\cite{Brownridge2011} A candidate for pathological water science is the autothixotropy of water - the observation that pure water will become more viscous after sitting still for a long time.\cite{Verdel2011} The reported autothixotropy effect meets two of Langmuir's key criteria for pathological science - the effect is at the threshold of detectability and is not consistently reproduced. Finally, the concept of ``water memory'' after high dilution has generated much pathological science. Although the first major experiment on water memory, whcih was published in \textit{Nature} in 1988,\cite{Davenas1988} has been thoroughly debunked,\cite{Maddox1988,Ernst2002} work continues to be published on water memory. Much of this research is supported by the lucrative homeopathy industry and published in a network of journals dedicated to the subject.

\section{Alternative explanations for EZ phenomena}
This section presents several alternative explanations to EZ phenomena - diffusiophoresis (long range chemotaxis), reported previously by Schurr, and van der Waals forces.  These theories provide quantitative explanations for the growth and maintenance of the exclusion zone where plastic microspheres made of (possibly functionalized) carboxylate, polystyrene, amidine, or polytetrafluoroethylene (PTFE) are repelled from various surfaces. 

\subsection{Diffusiophoresis}
\begin{figure*}[h]
\centering
\includegraphics[width=6cm]{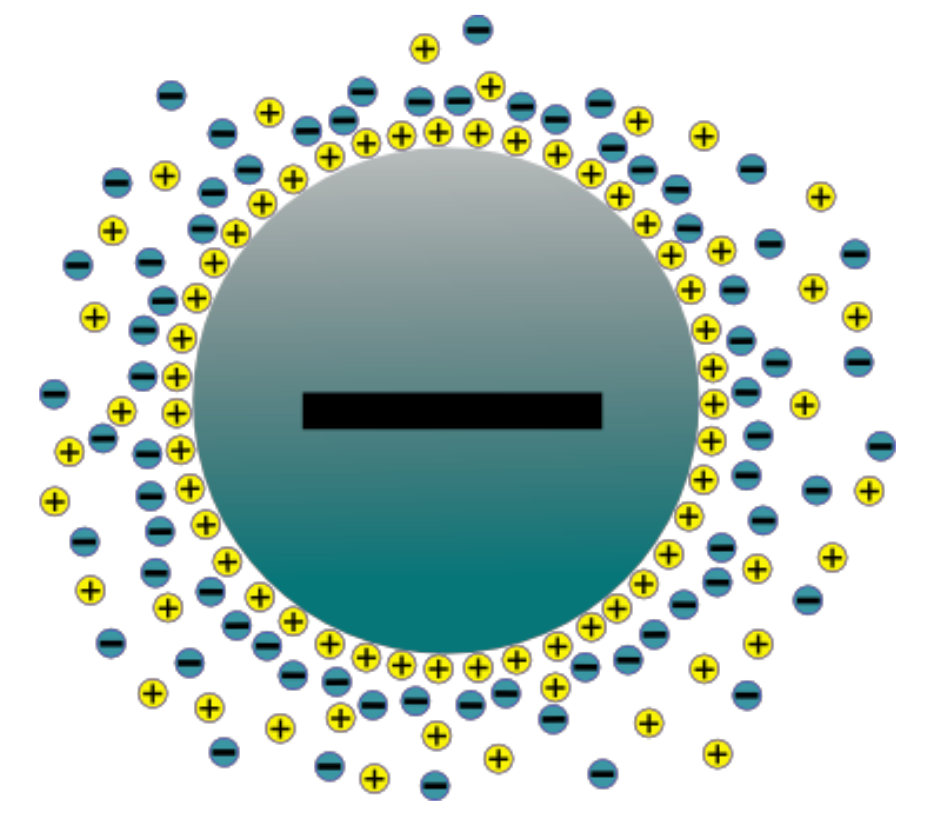}\quad\quad\quad\includegraphics[width=7.5cm]{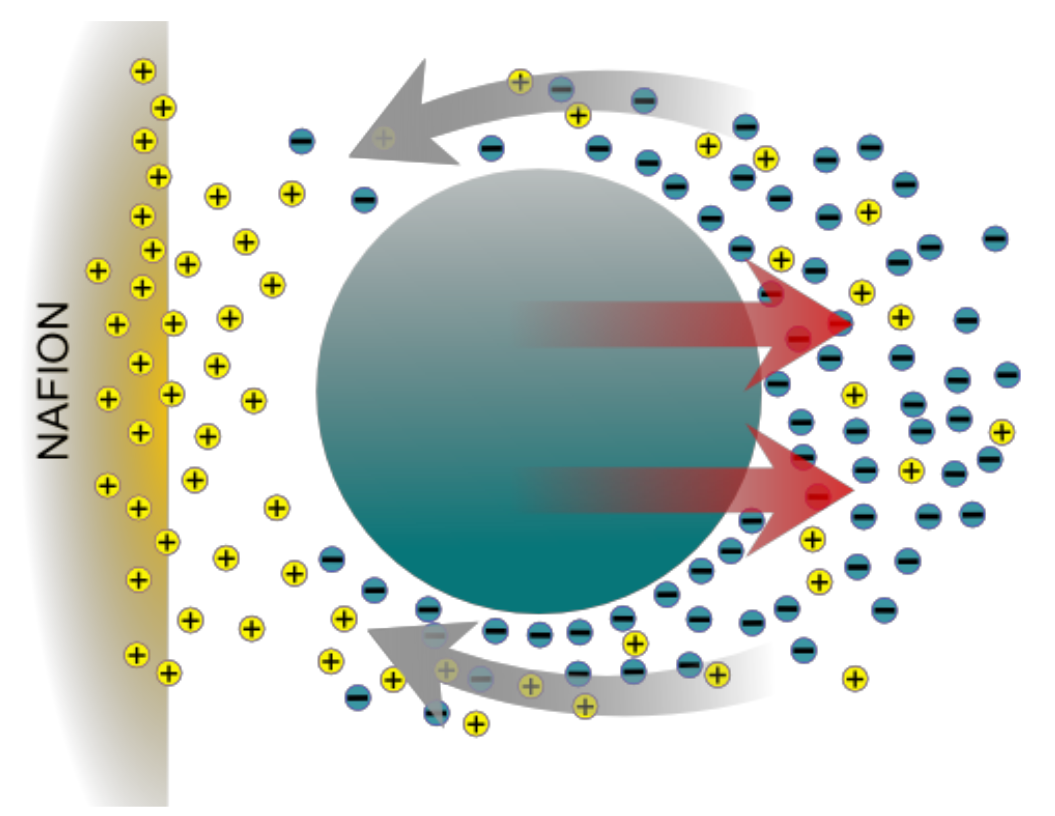}
\caption{(\textbf{left}) Homogeneous case.   (\textbf{right}) Heterogeneous case leading to diffusiophoresis.}
\label{fig1}
\end{figure*}   
Schurr (2013) has developed a theory which proposes that the EZ formation is created by forces arising from a concentration gradients of OH$^-$ or H$^+$ and salt. Called ``long range chemotaxis'' by Schurr,\cite{Schurr2013theory,Schurr2013} it is a type of a more general and well known phenomena in colloid science called diffusiophoresis. Huyghe, Wyss et al. (2014) propose that the EZs are generated by a combination of ion exchange and diffusiophoresis.\cite{Musa2013} They note that Nafion has an ample supply of exchangeable protons ready to exchange with cations in the solution. Such an exchange would create an inhomogeneous distribution of ions (salt gradient) in the liquid. According to the diffusiophoresis theory, a charged particle in an electrolyte solution would attract counter-ions (oppositely charged) via the influence of the local electric field. In a homogeneous solution it would be expected that the distribution of ions and counter-ions would be symmetrical around the particle. This would lead to a homogeneously distributed hydrostatic pressure with no fluid flow as shown in the left side of fig.~\ref{fig1}. However, with the introduction of a proton donor like Nafion the resulting inhomogeneous charge distribution would produce an asymmetrical arrangement of ions around the particle as shown in the right side of fig.~\ref{fig1}. In an effort to balance ions and counter ions a fluid flow results, propelling the particles away from the Nafion surface. 

Florea et al.\ have performed experiments on the EZ, carefully measuring its time course, and have shown that the data are fit by a model of diffusiophoresis.\cite{Florea2014} Notably, these experiments were done with the hydrophilic surface horizontal, which avoids convective fluid motions due to the force of gravity which occur when it is vertical, as in many of Pollack's experiments. Further experiments and a computational study using COMSOL Multiphysics simulation by Esplandiu et al.\ lend further support to the findings of Florea et al.\cite{Esplandiu2020}  Husz\'{a}r et al.\ note that the growth of the exclusion zone with times follows a power law with an exponent of 0.6, very close to the exponent of 0.5 expected for a diffusion-driven process.\cite{Huszr2014} Using laser tweezers, a forcefield has been measured inside the exclusion zone. Two independent experiments have found that the magnitude of the repulsive force decays as a function of distance from the surface in a manner consistent with the diffusiophoresis theory.\cite{Chen2011,Huszr2014} The presence of a force decaying from the surface is inconsistent with the theory that a new phase forms in the exclusion zone.

Pollack has responded to Shurr's original work.\cite{Pollack2013response} Figure 1 in Pollack's response arguably support the theory however, since it shows a large pH gradient, as indicated by a dye.\cite{Pollack2013response} However, in an earlier work Ovchinnikova \& Pollack argue that the pH gradients reflect storage and slow dissipation of electric charge by the EZ water rather than the Nafion.\cite{Ovchinnikova2009} 

Apart from the experiments mentioned previously, there are theoretical reasons to suppose that a large concentration gradient would arrise near the surface of Nafion, the most popular surface used for generating EZs. Nafion is a copolymer of tetrafluoroethylene and perfluoro-3,6-dioxa-4-methyl-7-octene sulfonic acid which finds application in fuel cell technology. If the sulfonic acid part were allowed to dissolve into water it would be quite a strong acid, but this doesn't happen since it remains bonded into the copolymer. When Nafion is placed in water it quickly swells, resulting in a gel-structure with an extremely high surface area. In this structure all of the sulphonic acid groups are surrounded by water. The highly negative sufonic acid group dissociates water and adsorbs $H+$ ions, resulting in a very low internal pH for Nafion, as observed with indicators such as methylene blue.\cite{Seger2007} Computational studies show it is energetically favorable for 2-4 hydronium ions to surround each sulfonic acid group.\cite{Phonyiem2011,Devanathan2010} Using methylene blue the internal acidity of Nafion has been estimated to be equivalent to 1.2M sulphuric acid.\cite{Seger2007} The excess protons inside Nafion are of two types - ``fixed'' ions which can ``hop'' between sulfonic groups, and ``mobile'' ions which can can freely diffuse away.\cite{Seger2007,Devanathan2010} Thus water around Nafion becomes acidic, with a pH gradient approaching neutral (7) further away from the memberane. This is shown clearly in experiments by Pollack where pH sensitive dyes have been added to the water.\cite{Chai2009} We have also observed this in our own experiments, were we also found that the average pH of the water around Nafion drops over the course of several days.\cite{Spencermastersthesis,Spencer2018}
 Elsewhere an acidic pH of water around Nafion has also been reported (pK$_a$ $\approx$ -6).\cite{Kreuer2000}

\subsection{EZs at metal surfaces: van der Waals repulsion and quantum phenomena}
The theory of chemotaxis of Schurr presents a compelling theory of the EZ phenomena observed near Nafion. However, Pollack's group has also reported EZ phenomena near metal surfaces, although they are much smaller in size.\cite{Chai2012metal} The EZ is largest for Zinc (220 $\mu$m), followed by aluminum, lead, tin, and tungsten (72 $\mu$m).\cite{Chai2012metal} Notably however, attempts to independently replicate these findings with aluminum and zinc have failed.\cite{Spencer2018} Pollack also reports EZ phenomena at the surface of platinum, but only after a voltage is applied.\cite{Musumeci2014}. While water molecules adsorb onto surfaces like platinum,\cite{Pedroza2015} and may dissociate on such surfaces in certain circumstances,\cite{Fajn2014} the expected gradient of hydronium ions as one moves away from the surface is expected to be small, if it exists at all. One possibility is that the exclusion zone phenomena near metals (and possibly other materials) may be partially explained by repulsive van der Waals forces (also called Casmir-Polder forces in this type of context). A role for van der Waals forces was first explored by De Ninno in 2017.\cite{DeNinno2017} In his calculation, he assumes that water contains coherent domains with higher dielectric constant and lower density (akin to the ``low density water phase'' in the famous two-phase model of water) and ``non-coherent domains'' with much lower dielectric constant ($\varepsilon \approx 13$). The existence of such domains has not been experimentally validated for room temperature water, and molecular dynamics simulations have not discovered such coherent structures (see \cite{Elton2014} and references therein).

The possibility that two objects of different composition may feel a repulsive force when submerged in a liquid was first realized by Hamaker in 1937.\cite{Hamaker1937} The full theory for such forces, for arbitrary dielectric media, was worked out by Lifshitz in 1954.\cite{Lifshitz:1956} Lifshitz’s equations allow for a repulsive force between two objects if the dielectric susceptibility of the medium between the two plates is intermediary between the two. Calculations using Lifshitz theory show that the finite size of the slabs does not effect the repulsion between them.\cite{vanZwol2010, Zhao2011}  Having free electrons, the dielectric constant of metals is extremely high (for instance Milling take the dielectric constant of gold to be 300).\cite{Milling1996} The dielectric constant of water is 78 and the dielectric constant of a polystyrene microsphere is about 2.5 (other plastic microspheres should have dielectric constants between 1.5 and 3). Thus, the metal-microsphere-water system obeys the conditions necessary for Casmir Pollard repulsion.

Most studies of the repulsive van der Waals force have used liquids other than water, likely due to the fact that water is easily contaminated with charge bearing solutes which can confound such experiments. The effect is also larger in nonpolar liquids than polar ones.\cite{Milling1996} Munday et al.\ (2009) have reported a repulsive Casmir force between a gold plate and a silica sphere submerged in bromobenzene.\cite{Munday2009} Similar repulsion has been found in follow up work with cyclohexane and other liquids.\cite{Meurk1997, Lee2002} Milling et al. (1996) measured the force between a gold sphere and PTFE block submerged in several liquids, including water.\cite{Milling1996} While their results for water were neutral/inconsistent (both weakly attractive and weakly repulsive forces were observed), their theoretical calculation indicates that the vdW force in water should be repulsive.\cite{Milling1996} There is clearly room for improvement in here, as the sign of the Hamaker constant predicted by theory only matches the sign found by experiment in 3/10 cases, suggesting an issue either with the theoretical calculation or experiments. Milling et al.\ note that the discrepancies most likely arise on the theoretical end as due to incomplete knowledge about the high frequency (UV) dielectric function of the materials involved, as the entire dielectric spectrum is required for the calculation.\cite{Milling1996}

One issue with this theory though is that retardation effects can diminish the van der Waals force starting at just a few nanometers of separation.\cite{Bevan1999,Lee2001} Retardation effects become important when the travel time due to the speed of light becomes similar the timescale (period) of polarization fluctuations which underlie the van der Waals force. Under retardation the force changes from falling as $1/r^7$ to  $1/r^8$. However, Isrealachvili notes that here is also a non-retarded zero frequency component to the vdW force which persists to large separations.\cite{Israelachvili2011} According to Isrealachvili, the actual progression of the vdW force may be from $1/r^7 \rightarrow 1/r^8 \rightarrow 1/r^7$.\cite{Israelachvili2011}

The growth of the EZ zone with laser light\cite{Chai2009} or MHz frequency electromagnetic fields,\cite{He2018} may be due to an induced van der Waals repulsion, although there may be a more prosaic explanation. It has been shown that the van der Waals forces between silver nanoparticles can be enhanced by radiation, since electromagnetic radiation induces fluctuating dipole moments in the particles. The possibility for light-driven enhancement of repulsive van der Waals forces has been shown theoretically by Rodr{\'{\i}}guez-Fortu{\~{n}}o et al.\cite{RodrguezFortuo2016} While these considerations are for metal nanoparticles, the polarizability of plastic (especially functionalized plastic) means such induced dipoles moments may be possible. Further theoretical study is needed to clarify this matter. 

\subsection{Other possible mechanisms and experimental confounds}
Husz\'{a}r et al.\ have investigated two other possible explanations for EZ-formation.\cite{Huszr2014}
\begin{itemize}
    \item Dissolution of Nafion, during which polymer strands diffusing out of the gel push the beads away from the surface.
    \item A ``brush mechanism'' in which closely spaced long elastic polymer strands keep the beads away by entropic forces.
\end{itemize}
Close inspection of gel showed that it does not loose mass, and an atomic force microscopy (AFM) study of the surface shows that there are no long strands hanging out, so they ruled out both of these mechanisms.

Bunkin et al.\ analyzed the swelling of Nafion with photo-luminescent UV spectroscopy, which detects the terminal sulfonic groups on Nafion fibers.\cite{Bunkin2018} Notably, they observed there are fibers which extend out into the liquid; for deuterium this matched roughly the exclusion zone distance observed in Pollack's experiments. The degree to which these fibers penetrate the liquid appears to be very sensitive to the presence of deuterium. As noted before, water-swollen Nafion absorbs UV strongly at 270 nm.\cite{Bunkin2019,DeAlmeida1997}

Apart from these two effects, there are other possible effects that can contaminate microsphere systems and confound experiments. Plastic nanospheres can be easily contaminated with charge bearing groups. In the case of PTFE these may include ``residual carboxylic groups from the polymerization process''.\cite{Milling1996} Referring to research that uses plastic microspheres Horinek et al.\ note  ``these systems are notoriously plagued by secondary effects, such as bubble adsorption and cavitation effects or compositional rearrangements''.\cite{Horinek2008} As an example, the discovery of an ultra-low frequency Debye relaxation in water, for instance, was later show to be due to microbubble contamination.\cite{Richert2011} There is also growing research showing that the removal of nanobubbles from water can be very challenging. This is especially true when they are adsorbed on surfaces. As noted before, the introduction of degassing methods reduced the appearance of the peak at 270 nm which Pollack attributes to EZ water.\cite{So2011} Thus careful degassing should be a key part of any research on EZ water going forward.

Finally, in passing we note that Chaplin has a theory which he calls ``self-generation of colligative properties''.\cite{2012arXivChaplin} The basic idea is an ``osmotic effect'' can be generated near hydrophilic surfaces, since the water molecules very close to the surface are moving slower, effectively resulting in lower temperature water near the surface. Chaplin predicts that an even larger osmotic effect should occur near nanobubble's surfaces, due to ``surface teathered'' solutes near or at the nanobubble air-water interface.\cite{chaplinweb} Chaplin's theory will require carefully designed experiments to test. 

\section{Conclusion}
In this review we noted several major problems with the theory that water in the EZ undergoes a phase change or significant reordering. We presented new results from neutron beam radiography which do not support the idea of a higher density phase and discussed how flaws were discovered in Pollack's birefringence measurements which have been suggested to support a structure change. Schurr's theory of macroscopic chemotaxis presents a compelling alternative theory which can explain experimental findings which Pollack's theory cannot, such as the precise time course of EZ growth, pH gradients emanating from the surface of Nafion, and the decaying forcefield measured by experiments with optical tweezers.\cite{Schurr2013theory,Florea2014,Esplandiu2020,Chen2011,Huszr2014} 
There are still many open questions about exclusion zones. The findings of EZs near different metal surfaces need to be better replicated and elaborated, as some attempts to replicate these findings have failed.\cite{Spencer2018} Many findings from Pollack's lab still need to be replicated by independent groups, in particular the growth of the EZ with laser irritation and the exclusion of salt. Both of these phenomena, if genuine, are in need of further explanation. Likewise, Rohani \& Pollack have observed anomalous flow in Nafion tubes, and understanding this phenomena may shed light on the ion dynamics around Nafion or here-to undiscovered experimental confounds.\cite{Rohani2013} A more complete understanding of the mechanisms behind EZ phenomena will assist in understanding their possible roles in biology as well as their possible engineering applications such as microfluidics and filtration.

\vspace{6pt}




\funding{This research received no external funding. The authors declare no conflict of interest. No National Institutes of Health funding or resources were used in the production of this work. One of the authors (Dr.\ Daniel C.\ Elton) wrote this article in his personal capacity. The opinions expressed in this article are the author's own and do not reflect the view of the National Institutes of Health, the Department of Health and Human Services, or the United States government. }


\conflictsofinterest{The author declares no conflict of interest.}



\reftitle{References}


\externalbibliography{yes}
\bibliography{bibliography}

\begin{thebibliography}{-------}
\providecommand{\natexlab}[1]{#1}

\bibitem[Chen \em{et~al.}(2011)Chen, Chung, Hsu, Wu, and Chin]{Chen2011}
Chen, C.S.; Chung, W.J.; Hsu, I.C.; Wu, C.M.; Chin, W.C.
\newblock Force field measurements within the exclusion zone of water.
\newblock {\em Journal of Biological Physics} {\bf 2011}, {\em 38},~113--120.

\bibitem[Musa \em{et~al.}(2013)Musa, Florea, van Loon, Wyss, and
  Huyghe]{Musa2013}
Musa, S.; Florea, D.; van Loon, S.; Wyss, H.; Huyghe, J.M.
\newblock Interfacial Water: Unexplained Phenomena.
\newblock  Poromechanics V. American Society of Civil Engineers,  2013.

\bibitem[Husz{\'{a}}r \em{et~al.}(2014)Husz{\'{a}}r, M{\'{a}}rtonfalvi, Laki,
  Iv{\'{a}}n, and Kellermayer]{Huszr2014}
Husz{\'{a}}r, I.; M{\'{a}}rtonfalvi, Z.; Laki, A.; Iv{\'{a}}n, K.; Kellermayer,
  M.
\newblock Exclusion-Zone Dynamics Explored with Microfluidics and Optical
  Tweezers.
\newblock {\em Entropy} {\bf 2014}, {\em 16},~4322--4337.

\bibitem[Gudkov \em{et~al.}(2014)Gudkov, Astashev, Bruskov, Kozlov, Zakharov,
  and Bunkin]{Gudkov2014}
Gudkov, S.; Astashev, M.; Bruskov, V.; Kozlov, V.; Zakharov, S.; Bunkin, N.
\newblock Self-oscillating Water Chemiluminescence Modes and Reactive Oxygen
  Species Generation Induced by Laser Irradiation; Effect of the Exclusion Zone
  Created by Nafion.
\newblock {\em Entropy} {\bf 2014}, {\em 16},~6166--6185.

\bibitem[Jabs and Rubik(2014)]{Jabs2014}
Jabs, H.; Rubik, B.
\newblock Self-Organization at Aqueous Colloid-Membrane Interfaces and an
  Optical Method to Measure the Kinetics of Exclusion Zone Formation.
\newblock {\em Entropy} {\bf 2014}, {\em 16},~5954--5975.

\bibitem[Florea \em{et~al.}(2014)Florea, Musa, Huyghe, and Wyss]{Florea2014}
Florea, D.; Musa, S.; Huyghe, J.M.R.; Wyss, H.M.
\newblock Long-range repulsion of colloids driven by ion exchange and
  diffusiophoresis.
\newblock {\em Proceedings of the National Academy of Sciences} {\bf 2014},
  {\em 111},~6554--6559.

\bibitem[Bunkin \em{et~al.}(2013)Bunkin, Ignatiev, Kozlov, Shkirin, Zakharov,
  and Zinchenko]{bunkin2013}
Bunkin, N.F.; Ignatiev, P.S.; Kozlov, V.A.; Shkirin, A.V.; Zakharov, S.D.;
  Zinchenko, A.A.
\newblock Study of the Phase States of Water Close to Nafion Interface.
\newblock {\em WATER Journal} {\bf 2013}, {\em 4}.

\bibitem[Yakhno and Yakhno(2018)]{Yakhno2018}
Yakhno, T.A.; Yakhno, V.G.
\newblock On the interaction of water with hydrophilic surfaces.
\newblock {\em Russ. J. Biol. Phys. Chem.} {\bf 2018}, {\em 3},~9–18.

\bibitem[Spencer \em{et~al.}(2018)Spencer, Riches, and Williams]{Spencer2018}
Spencer, P.D.; Riches, J.D.; Williams, E.D.
\newblock Exclusion zone water is associated with material that exhibits proton
  diffusion but not birefringent properties.
\newblock {\em Fluid Phase Equilibria} {\bf 2018}, {\em 466},~103--109.

\bibitem[Sharma \em{et~al.}(2018)Sharma, Adams, Cashdollar, Li, Nguyen, Sai,
  Shi, Velchuru, Zhu, and Pollack]{Sharma2018}
Sharma, A.; Adams, C.; Cashdollar, B.D.; Li, Z.; Nguyen, N.V.; Sai, H.; Shi,
  J.; Velchuru, G.; Zhu, K.Z.; Pollack, G.H.
\newblock Effect of Health-Promoting Agents on Exclusion-Zone Size.
\newblock {\em Dose-Response} {\bf 2018}, {\em 16},~155932581879693.

\bibitem[Esplandiu \em{et~al.}(2020)Esplandiu, Reguera, and
  Fraxedas]{Esplandiu2020}
Esplandiu, M.J.; Reguera, D.; Fraxedas, J.
\newblock Electrophoretic origin of long-range repulsion of colloids near
  water/{Nafion} interfaces.
\newblock {\em Soft Matter} {\bf 2020}, {\em 16},~3717--3726.

\bibitem[Zheng \em{et~al.}(2006)Zheng, Chin, Khijniak, Khijniak, and
  Pollack]{Zheng2006EZ}
Zheng, J.; Chin, W.; Khijniak, E.; Khijniak, E.; Pollack, G.H.
\newblock Surfaces and interfacial water: Evidence that hydrophilic surfaces
  have long-range impact.
\newblock {\em Advances in Colloid and Interface Science} {\bf 2006}, {\em
  127},~19--27.

\bibitem[Pedroza \em{et~al.}(2015)Pedroza, Poissier, and
  Fern{\'{a}}ndez-Serra]{Pedroza2015}
Pedroza, L.S.; Poissier, A.; Fern{\'{a}}ndez-Serra, M.V.
\newblock Local order of liquid water at metallic electrode surfaces.
\newblock {\em The Journal of Chemical Physics} {\bf 2015}, {\em 142},~034706.

\bibitem[Chai \em{et~al.}(2012)Chai, Mahtani, and Pollack]{Chai2012metal}
Chai, B.; Mahtani, A.G.; Pollack, G.H.
\newblock Unexpected presence of solute-free zones at metal-water interfaces.
\newblock {\em Contemporary Materials} {\bf 2012}, {\em 1}.

\bibitem[Sulbar{\'{a}}n \em{et~al.}(2014)Sulbar{\'{a}}n, Toriz, Allan, Pollack,
  and Delgado]{Sulbarn2014}
Sulbar{\'{a}}n, B.; Toriz, G.; Allan, G.G.; Pollack, G.H.; Delgado, E.
\newblock The dynamic development of exclusion zones on cellulosic surfaces.
\newblock {\em Cellulose} {\bf 2014}, {\em 21},~1143--1148.
\newblock
  doi:{\changeurlcolor{black}\href{https://doi.org/10.1007/s10570-014-0165-y}{\detokenize{10.1007/s10570-014-0165-y}}}.

\bibitem[Pedregal-Cort{\'{e}}s \em{et~al.}(2019)Pedregal-Cort{\'{e}}s, Toriz,
  Delgado, and Pollack]{PedregalCorts2019}
Pedregal-Cort{\'{e}}s, R.; Toriz, G.; Delgado, E.; Pollack, G.H.
\newblock Interfacial water and its potential role in the function of sericin
  against biofouling.
\newblock {\em Biofouling} {\bf 2019}, {\em 35},~732--741.
\newblock
  doi:{\changeurlcolor{black}\href{https://doi.org/10.1080/08927014.2019.1653863}{\detokenize{10.1080/08927014.2019.1653863}}}.

\bibitem[Ball(2008)]{Ball2008}
Ball, P.
\newblock Water as an Active Constituent in Cell Biology.
\newblock {\em Chemical Reviews} {\bf 2008}, {\em 108},~74--108.

\bibitem[Drost-Hansen(1969)]{DrostHansen1969}
Drost-Hansen, W.
\newblock Structure of Water Near Solid Interfaces.
\newblock {\em Industrial {\&} Engineering Chemistry} {\bf 1969}, {\em
  61},~10--47.

\bibitem[Drost-Hansen(1973)]{DrostHansen1973}
Drost-Hansen, W.
\newblock Phase transitions in biological systems: manifestations of
  cooperative processes in vicinal water.
\newblock {\em Annals of the New York Academy of Sciences} {\bf 1973}, {\em
  204},~100--112.

\bibitem[Fenter and Sturchio(2004)]{Fenter2004}
Fenter, P.; Sturchio, N.C.
\newblock Mineral{\textendash}water interfacial structures revealed by
  synchrotron X-ray scattering.
\newblock {\em Progress in Surface Science} {\bf 2004}, {\em 77},~171--258.

\bibitem[Verdaguer \em{et~al.}(2006)Verdaguer, Sacha, Bluhm, and
  Salmeron]{Verdaguer2006}
Verdaguer, A.; Sacha, G.M.; Bluhm, H.; Salmeron, M.
\newblock Molecular Structure of Water at Interfaces:~ Wetting at the Nanometer
  Scale.
\newblock {\em Chemical Reviews} {\bf 2006}, {\em 106},~1478--1510.

\bibitem[Maccarini(2007)]{Maccarini2007}
Maccarini, M.
\newblock Water at solid surfaces: A review of selected theoretical aspects and
  experiments on the subject.
\newblock {\em Biointerphases} {\bf 2007}, {\em 2},~MR1--MR15.

\bibitem[Ling(2003)]{Ling2003}
Ling, G.N.
\newblock A new theoretical foundation for the polarized-oriented multilayer
  theory of cell water and for inanimate systems demonstrating long-range
  dynamic structuring of water molecules.
\newblock {\em Physiological chemistry and physics and medical NMR} {\bf 2003},
  {\em 35},~91—130.

\bibitem[Deryagin~BV(1986)]{Deragin1986}
Deryagin~BV, G.M.
\newblock Electromagnetic nature of forces of repulsion forming aureoles around
  cells.
\newblock {\em Colloid J. USSR} {\bf 1986}, {\em 48},~209–211.

\bibitem[Shen and Ostroverkhov(2006)]{Shen2006}
Shen, Y.R.; Ostroverkhov, V.
\newblock Sum-Frequency Vibrational Spectroscopy on Water Interfaces:~ Polar
  Orientation of Water Molecules at Interfaces.
\newblock {\em Chemical Reviews} {\bf 2006}, {\em 106},~1140--1154.

\bibitem[Catalano(2011)]{Catalano2011}
Catalano, J.G.
\newblock Weak interfacial water ordering on isostructural hematite and
  corundum (001) surfaces.
\newblock {\em Geochimica et Cosmochimica Acta} {\bf 2011}, {\em
  75},~2062--2071.

\bibitem[Eftekhari-Bafrooei and Borguet(2010)]{EftekhariBafrooei2010}
Eftekhari-Bafrooei, A.; Borguet, E.
\newblock Effect of Hydrogen-Bond Strength on the Vibrational Relaxation of
  Interfacial Water.
\newblock {\em Journal of the American Chemical Society} {\bf 2010}, {\em
  132},~3756--3761.

\bibitem[Ebbinghaus \em{et~al.}(2007)Ebbinghaus, Kim, Heyden, Yu, Heugen,
  Gruebele, Leitner, and Havenith]{Ebbinghaus2007}
Ebbinghaus, S.; Kim, S.J.; Heyden, M.; Yu, X.; Heugen, U.; Gruebele, M.;
  Leitner, D.M.; Havenith, M.
\newblock An extended dynamical hydration shell around proteins.
\newblock {\em Proceedings of the National Academy of Sciences} {\bf 2007},
  {\em 104},~20749--20752.

\bibitem[Elton and Fern{\'{a}}ndez-Serra(2014)]{Elton2014}
Elton, D.C.; Fern{\'{a}}ndez-Serra, M.V.
\newblock Polar nanoregions in water: A study of the dielectric properties of
  {TIP}4P/2005, {TIP}4P/2005f and {TTM}3F.
\newblock {\em The Journal of Chemical Physics} {\bf 2014}, {\em 140},~124504.

\bibitem[Elton(2016)]{EltonThesis}
Elton, D.C.
\newblock PhD Thesis : ``Understanding the Dielectric Properties of Water''.
\newblock PhD thesis, Stony Brook University,  2016.

\bibitem[Chaplin(2000)]{Chaplin2000}
Chaplin, M.
\newblock A proposal for the structuring of water.
\newblock {\em Biophysical Chemistry} {\bf 2000}, {\em 83},~211--221.

\bibitem[Park \em{et~al.}(2006)Park, Choi, Woo, and Moon]{Park2006}
Park, J.S.; Choi, J.H.; Woo, J.J.; Moon, S.H.
\newblock An electrical impedance spectroscopic ({EIS}) study on transport
  characteristics of ion-exchange membrane systems.
\newblock {\em Journal of Colloid and Interface Science} {\bf 2006}, {\em
  300},~655--662.

\bibitem[ming Zheng \em{et~al.}(2009)ming Zheng, Wexler, and
  Pollack]{Zheng2009ordered}
ming Zheng, J.; Wexler, A.; Pollack, G.H.
\newblock Effect of buffers on aqueous solute-exclusion zones around
  ion-exchange resins.
\newblock {\em Journal of Colloid and Interface Science} {\bf 2009}, {\em
  332},~511--514.
\newblock
  doi:{\changeurlcolor{black}\href{https://doi.org/10.1016/j.jcis.2009.01.010}{\detokenize{10.1016/j.jcis.2009.01.010}}}.

\bibitem[Klimov and Pollack(2007)]{Klimov2007}
Klimov, A.; Pollack, G.H.
\newblock Visualization of Charge-Carrier Propagation in Water.
\newblock {\em Langmuir} {\bf 2007}, {\em 23},~11890--11895.

\bibitem[ming Zheng and Pollack(2003)]{Zheng2003}
ming Zheng, J.; Pollack, G.H.
\newblock Long-range forces extending from polymer-gel surfaces.
\newblock {\em Physical Review E} {\bf 2003}, {\em 68}.

\bibitem[Chai \em{et~al.}(2009)Chai, Yoo, and Pollack]{Chai2009}
Chai, B.; Yoo, H.; Pollack, G.H.
\newblock Effect of Radiant Energy on Near-Surface Water.
\newblock {\em The Journal of Physical Chemistry B} {\bf 2009}, {\em
  113},~13953--13958.

\bibitem[Chai \em{et~al.}(2012)Chai, Mahtani, and Pollack]{Chai20122}
Chai, B.; Mahtani, A.G.; Pollack, G.H.
\newblock Unexpected presence of solute-free zones at metal-water interfaces.
\newblock {\em Contemporary Materials} {\bf 2012}, {\em 1}.

\bibitem[Pollack(2013)]{pollack2013fourth}
Pollack, G.
\newblock {\em The Fourth Phase of Water: Beyond Solid, Liquid, and Vapor};
  Ebner \& Sons,  2013.

\bibitem[Hwang \em{et~al.}(2018)Hwang, Hong, Sharma, Pollack, and
  Bahng]{Hwang2018}
Hwang, S.G.; Hong, J.K.; Sharma, A.; Pollack, G.H.; Bahng, G.
\newblock Exclusion zone and heterogeneous water structure at ambient
  temperature.
\newblock {\em {PLOS} {ONE}} {\bf 2018}, {\em 13},~e0195057.

\bibitem[Y \em{et~al.}(2015)Y, Takizawa, and Lohwacharin]{ZhangWATER2015}
Y, Y.Z.; Takizawa, S.; Lohwacharin, J.
\newblock Spontaneous Particle Separation and Salt Rejection by Hydrophilic
  Membranes.
\newblock {\em Water Journal} {\bf 2015}, {\em 7}.
\newblock
  doi:{\changeurlcolor{black}\href{https://doi.org/10.14294/water.2015.2}{\detokenize{10.14294/water.2015.2}}}.

\bibitem[Figueroa and Pollack(2011)]{Figueroa2011}
Figueroa, X.A.; Pollack, G.H.
\newblock Exclusion-zone formation from discontinuous nafion surfaces.
\newblock {\em International Journal of Design {\&} Nature and Ecodynamics}
  {\bf 2011}, {\em 6},~286--296.
\newblock
  doi:{\changeurlcolor{black}\href{https://doi.org/10.2495/dne-v6-n4-286-296}{\detokenize{10.2495/dne-v6-n4-286-296}}}.

\bibitem[Oehr and LeMay(2014)]{Oehr2014}
Oehr, K.; LeMay, P.
\newblock The Case for Tetrahedral Oxy-subhydride ({TOSH}) Structures in the
  Exclusion Zones of Anchored Polar Solvents Including Water.
\newblock {\em Entropy} {\bf 2014}, {\em 16},~5712--5720.

\bibitem[Giudice \em{et~al.}(2013)Giudice, Tedeschi, Vitiello, and
  Voeikov]{Giudice2013}
Giudice, E.D.; Tedeschi, A.; Vitiello, G.; Voeikov, V.
\newblock Coherent structures in liquid water close to hydrophilic surfaces.
\newblock {\em Journal of Physics: Conference Series} {\bf 2013}, {\em
  442},~012028.
\newblock
  doi:{\changeurlcolor{black}\href{https://doi.org/10.1088/1742-6596/442/1/012028}{\detokenize{10.1088/1742-6596/442/1/012028}}}.

\bibitem[Elia \em{et~al.}(2019)Elia, Napoli, Germano, Oliva, Roviello, Niccoli,
  Amoresano, Naviglio, Ciaravolo, Trifuoggi, and Yinnon]{Elia2019}
Elia, V.; Napoli, E.; Germano, R.; Oliva, R.; Roviello, V.; Niccoli, M.;
  Amoresano, A.; Naviglio, D.; Ciaravolo, M.; Trifuoggi, M.; Yinnon, T.A.
\newblock New chemical-physical properties of water after iterative procedure
  using hydrophilic polymers: The case of paper filter.
\newblock {\em Journal of Molecular Liquids} {\bf 2019}, {\em 296},~111808.

\bibitem[Giudice \em{et~al.}(2010)Giudice, Spinetti, and Tedeschi]{Giudice2010}
Giudice, E.D.; Spinetti, P.R.; Tedeschi, A.
\newblock Water Dynamics at the Root of Metamorphosis in Living Organisms.
\newblock {\em Water} {\bf 2010}, {\em 2},~566--586.
\newblock
  doi:{\changeurlcolor{black}\href{https://doi.org/10.3390/w2030566}{\detokenize{10.3390/w2030566}}}.

\bibitem[So \em{et~al.}(2011)So, Stahlberg, and Pollack]{So2011}
So, E.; Stahlberg, R.; Pollack, G.H.
\newblock Exclusion zone as intermediate between ice and water.
\newblock  Water and Society. {WIT} Press,  2011.

\bibitem[Fedyakin(1962)]{Fedyakin1962}
Fedyakin, N.N.
\newblock Change in the structure of water during condensation in capillaries.
\newblock {\em Kolloid Zh.} {\bf 1962}, {\em 24}.

\bibitem[Hasted(1971)]{Hasted1971}
Hasted, J.B.
\newblock Water and ``polywater''.
\newblock {\em Contemporary Physics} {\bf 1971}, {\em 12},~133--152.

\bibitem[Segarra-Mart{\'{\i}}
  \em{et~al.}(2014{\natexlab{a}})Segarra-Mart{\'{\i}}, Roca-Sanju{\'{a}}n, and
  Merch{\'{a}}n]{SegarraMart2014}
Segarra-Mart{\'{\i}}, J.; Roca-Sanju{\'{a}}n, D.; Merch{\'{a}}n, M.
\newblock Can the Hexagonal Ice-like Model Render the Spectroscopic
  Fingerprints of Structured Water? Feedback from Quantum-Chemical
  Computations.
\newblock {\em Entropy} {\bf 2014}, {\em 16},~4101--4120.

\bibitem[Segarra-Mart{\'{\i}}
  \em{et~al.}(2014{\natexlab{b}})Segarra-Mart{\'{\i}}, Roca-Sanju{\'{a}}n, and
  Merch{\'{a}}n]{SegarraMart2014MDPI}
Segarra-Mart{\'{\i}}, J.; Roca-Sanju{\'{a}}n, D.; Merch{\'{a}}n, M.
\newblock Can the Hexagonal Ice-like Model Render the Spectroscopic
  Fingerprints of Structured Water? Feedback from Quantum-Chemical
  Computations.
\newblock {\em Entropy} {\bf 2014}, {\em 16},~4101--4120.

\bibitem[Giudice \em{et~al.}(2015)Giudice, Voeikov, Tedeschi, and
  Vitiello]{Giudicebook}
Giudice, E.; Voeikov, V.; Tedeschi, A.; Vitiello, G.
\newblock {\em The origin and the special role of coherent water in living
  systems}; Vol. 37661,  2015; pp. 95--111.
\newblock
  doi:{\changeurlcolor{black}\href{https://doi.org/10.13140/RG.2.1.2329.1046}{\detokenize{10.13140/RG.2.1.2329.1046}}}.

\bibitem[Chai and Pollack(2010)]{Chai2010}
Chai, B.; Pollack, G.H.
\newblock Solute-Free Interfacial Zones in Polar Liquids.
\newblock {\em The Journal of Physical Chemistry B} {\bf 2010}, {\em
  114},~5371--5375.

\bibitem[Skinner \em{et~al.}(2012)Skinner, Benmore, Shyam, Weber, and
  Parise]{Skinner2012}
Skinner, L.B.; Benmore, C.J.; Shyam, B.; Weber, J.K.R.; Parise, J.B.
\newblock Structure of the floating water bridge and water in an electric
  field.
\newblock {\em Proceedings of the National Academy of Sciences} {\bf 2012},
  {\em 109},~16463--16468.

\bibitem[Fuchs \em{et~al.}(2009)Fuchs, Bitschnau, Woisetschl\"{a}ger, Maier,
  Beuneu, and Teixeira]{Fuchs2009neutron}
Fuchs, E.C.; Bitschnau, B.; Woisetschl\"{a}ger, J.; Maier, E.; Beuneu, B.;
  Teixeira, J.
\newblock Neutron scattering of a floating heavy water bridge.
\newblock {\em Journal of Physics D: Applied Physics} {\bf 2009}, {\em
  42},~065502.

\bibitem[Chai \em{et~al.}(2008)Chai, Zheng, Zhao, and
  Pollack]{Chai2008absorption}
Chai, B.; Zheng, J.; Zhao, Q.; Pollack, G.H.
\newblock Spectroscopic Studies of Solutes in Aqueous Solution.
\newblock {\em The Journal of Physical Chemistry A} {\bf 2008}, {\em
  112},~2242--2247.

\bibitem[Dibble \em{et~al.}(2014)Dibble, J, and Tiller]{DibleWATER2014}
Dibble, W.E.; J, J.K.; Tiller, W.A.
\newblock Bulk Water with Exclusion Zone Water Characteristics: Experimental
  Evidence of Interaction With a Non-physical Agent.
\newblock {\em Water Journal} {\bf 2014}, {\em 6}.
\newblock
  doi:{\changeurlcolor{black}\href{https://doi.org/10.14294/water.2013.14}{\detokenize{10.14294/water.2013.14}}}.

\bibitem[Bielski \em{et~al.}(1985)Bielski, Cabelli, Arudi, and
  Ross]{Bielski1985}
Bielski, B.H.J.; Cabelli, D.E.; Arudi, R.L.; Ross, A.B.
\newblock Reactivity of {HO}2/O-2 Radicals in Aqueous Solution.
\newblock {\em Journal of Physical and Chemical Reference Data} {\bf 1985},
  {\em 14},~1041--1100.
\newblock
  doi:{\changeurlcolor{black}\href{https://doi.org/10.1063/1.555739}{\detokenize{10.1063/1.555739}}}.

\bibitem[Janik and Tripathi(2013)]{Janik2013}
Janik, I.; Tripathi, G.N.R.
\newblock The nature of the superoxide radical anion in water.
\newblock {\em The Journal of Chemical Physics} {\bf 2013}, {\em 139},~014302.
\newblock
  doi:{\changeurlcolor{black}\href{https://doi.org/10.1063/1.4811697}{\detokenize{10.1063/1.4811697}}}.

\bibitem[Matheson and Lee(1969)]{Matheson1969}
Matheson, I.B.C.; Lee, J.
\newblock The Absorption Spectrum of Superoxide Anion in Dimethylsulfoxide.
\newblock {\em Spectroscopy Letters} {\bf 1969}, {\em 2},~117--119.
\newblock
  doi:{\changeurlcolor{black}\href{https://doi.org/10.1080/00387016908050026}{\detokenize{10.1080/00387016908050026}}}.

\bibitem[Czapski and Dorfman(1964)]{Czapski1964}
Czapski, G.; Dorfman, L.M.
\newblock Pulse Radiolysis Studies. V. Transient Spectra and Rate Constants in
  Oxygenated Aqueous Solutions1.
\newblock {\em The Journal of Physical Chemistry} {\bf 1964}, {\em
  68},~1169--1177.
\newblock
  doi:{\changeurlcolor{black}\href{https://doi.org/10.1021/j100787a034}{\detokenize{10.1021/j100787a034}}}.

\bibitem[Shin()]{shinNMRweb}
Shin, P.
\newblock Water, Water, Everywhere, Caveat Emptor (Buyer Beware)!
\newblock \url{http://www.csun.edu/~alchemy/Caveat_Emptor.pdf}.
\newblock Accessed: 2020-05-09.

\bibitem[Pokorn{\'{y}} \em{et~al.}(2015)Pokorn{\'{y}}, Pokorn{\'{y}}, Foletti,
  Kobilkov{\'{a}}, Vrba, and Vrba]{Pokorn2015:675}
Pokorn{\'{y}}, J.; Pokorn{\'{y}}, J.; Foletti, A.; Kobilkov{\'{a}}, J.; Vrba,
  J.; Vrba, J.
\newblock Mitochondrial Dysfunction and Disturbed Coherence: Gate to Cancer.
\newblock {\em Pharmaceuticals} {\bf 2015}, {\em 8},~675--695.

\bibitem[Hwang \em{et~al.}(2017)Hwang, Lee, Lee, and Bahng]{Hwang2017}
Hwang, S.G.; Lee, H.S.; Lee, B.C.; Bahng, G.
\newblock Effect of Antioxidant Water on the Bioactivities of Cells.
\newblock {\em International Journal of Cell Biology} {\bf 2017}, {\em
  2017},~1--12.

\bibitem[Tychinsky(2014)]{Tychinsky2014}
Tychinsky, V.P.
\newblock Extension of the concept of an anomalous water component to images of
  T-cell organelles.
\newblock {\em Journal of Biomedical Optics} {\bf 2014}, {\em 19},~126008.

\bibitem[Spencer(2018)]{Spencermastersthesis}
Spencer, P.
\newblock Examining claims of long-range molecular order in water molecules.
\newblock Master's thesis, Queensland University of Technology,  2018.

\bibitem[Takenaka \em{et~al.}(2013)Takenaka, Sugimoto, Takami, Sugioka,
  Tsukada, Adschiri, and Saito]{Takenaka2013}
Takenaka, N.; Sugimoto, K.; Takami, S.; Sugioka, K.; Tsukada, T.; Adschiri, T.;
  Saito, Y.
\newblock Application of Neutron Radiography to Flow Visualization in
  Supercritical Water.
\newblock {\em Physics Procedia} {\bf 2013}, {\em 43},~264--268.

\bibitem[Das and Pollack(2013)]{Das2013}
Das, R.; Pollack, G.H.
\newblock Charge-Based Forces at the Nafion{\textendash}Water Interface.
\newblock {\em Langmuir} {\bf 2013}, {\em 29},~2651--2658.

\bibitem[Totland and Nerdal(2016)]{Totland2016}
Totland, C.; Nerdal, W.
\newblock Experimental Determination of Water Molecular Orientation near a
  Silica Surface Using {NMR} Spectroscopy.
\newblock {\em The Journal of Physical Chemistry C} {\bf 2016}, {\em
  120},~5052--5058.
\newblock
  doi:{\changeurlcolor{black}\href{https://doi.org/10.1021/acs.jpcc.6b00466}{\detokenize{10.1021/acs.jpcc.6b00466}}}.

\bibitem[Yoo \em{et~al.}(2016)Yoo, Baker, Pirie, Hovakeemian, and
  Pollack]{Yoobihan2016water}
Yoo, H.; Baker, D.R.; Pirie, C.M.; Hovakeemian, B.; Pollack, G.H.,
  Characteristics of water adjacent to hydrophilic interfaces.
\newblock In {\em Water: The Forgotten Biological Molecule}; Pan Stanford
  Publishing,  2016; p. 123.

\bibitem[Tychinsky(2011)]{Tychinsky2011}
Tychinsky, V.
\newblock High Electric Susceptibility is the Signature of Structured Water in
  Water-Containing Objects.
\newblock {\em Water Journal} {\bf 2011}, {\em 3}.

\bibitem[Shokr and Sinha(2015)]{icelab2015}
Shokr, M.; Sinha, N.
\newblock Laboratory Techniques for Revealing the Structure of Polycrystalline
  Ice. In {\em Sea Ice}; John Wiley \&, Sons, Inc,  2015; pp. 231--269.

\bibitem[Deragin \em{et~al.}(1967)Deragin, Churaev, Fedyakin, Talaev, and
  Ershova]{Deragin1967}
Deragin, B.V.; Churaev, N.V.; Fedyakin, N.N.; Talaev, M.V.; Ershova, I.G.
\newblock The modified state of water and other liquids.
\newblock {\em Bulletin of the Academy of Sciences of the {USSR} Division of
  Chemical Science} {\bf 1967}, {\em 16},~2095--2102.

\bibitem[Lippincott \em{et~al.}(1969)Lippincott, Stromberg, Grant, and
  Cessac]{Lippincott1969AAAS}
Lippincott, E.R.; Stromberg, R.R.; Grant, W.H.; Cessac, G.L.
\newblock Polywater.
\newblock {\em Science} {\bf 1969}, {\em 164},~1482--1487.

\bibitem[Eisenberg(1981)]{EISENBERG1981Science}
Eisenberg, D.
\newblock A Scientific Gold Rush.
\newblock {\em Science} {\bf 1981}, {\em 213},~1104--1105.

\bibitem[Rousseau and Porto(1970)]{Rousseau1970Science}
Rousseau, D.L.; Porto, S.P.S.
\newblock Polywater: Polymer or Artifact?
\newblock {\em Science} {\bf 1970}, {\em 167},~1715--1719.

\bibitem[Rousseau(1971)]{Rousseau1971Science}
Rousseau, D.L.
\newblock ``Polywater'' and Sweat: Similarities between the Infrared Spectra.
\newblock {\em Science} {\bf 1971}, {\em 171},~170--172.

\bibitem[Bennion and Neuton(1976)]{Bennion1976}
Bennion, B.C.; Neuton, L.A.
\newblock The Epidemiology of Research on {\textquotedblleft}Anomalous
  Water{\textquotedblright}.
\newblock {\em Journal of the American Society for Information Science} {\bf
  1976}, {\em 27},~53--56.

\bibitem[Brownridge(2011)]{Brownridge2011}
Brownridge, J.D.
\newblock When does hot water freeze faster then cold water? A search for the
  Mpemba effect.
\newblock {\em American Journal of Physics} {\bf 2011}, {\em 79},~78--84.

\bibitem[Verdel \em{et~al.}(2011)Verdel, Jerman, and Bukovec]{Verdel2011}
Verdel, N.; Jerman, I.; Bukovec, P.
\newblock The {\textquotedblleft}Autothixotropic{\textquotedblright} Phenomenon
  of Water and its Role in Proton Transfer.
\newblock {\em International Journal of Molecular Sciences} {\bf 2011}, {\em
  12},~7481--7494.

\bibitem[Davenas \em{et~al.}(1988)Davenas, Beauvais, Amara, Oberbaum, Robinzon,
  Miadonnai, Tedeschi, Pomeranz, Fortner, Belon, Sainte-Laudy, Poitevin, and
  Benveniste]{Davenas1988}
Davenas, E.; Beauvais, F.; Amara, J.; Oberbaum, M.; Robinzon, B.; Miadonnai,
  A.; Tedeschi, A.; Pomeranz, B.; Fortner, P.; Belon, P.; Sainte-Laudy, J.;
  Poitevin, B.; Benveniste, J.
\newblock Human basophil degranulation triggered by very dilute antiserum
  against {IgE}.
\newblock {\em Nature} {\bf 1988}, {\em 333},~816--818.
\newblock
  doi:{\changeurlcolor{black}\href{https://doi.org/10.1038/333816a0}{\detokenize{10.1038/333816a0}}}.

\bibitem[Maddox \em{et~al.}(1988)Maddox, Randi, and Stewart]{Maddox1988}
Maddox, J.; Randi, J.; Stewart, W.W.
\newblock "High-dilution" experiments a delusion.
\newblock {\em Nature} {\bf 1988}, {\em 334},~287--290.

\bibitem[Ernst(2002)]{Ernst2002}
Ernst, E.
\newblock A systematic review of systematic reviews of homeopathy.
\newblock {\em British Journal of Clinical Pharmacology} {\bf 2002}, {\em
  54},~577--582.

\bibitem[Schurr \em{et~al.}(2013)Schurr, Fujimoto, Huynh, and
  Chiu]{Schurr2013theory}
Schurr, J.M.; Fujimoto, B.S.; Huynh, L.; Chiu, D.T.
\newblock A Theory of Macromolecular Chemotaxis.
\newblock {\em The Journal of Physical Chemistry B} {\bf 2013}, {\em
  117},~7626--7652.

\bibitem[Schurr(2013)]{Schurr2013}
Schurr, J.M.
\newblock Phenomena Associated with Gel{\textendash}Water Interfaces. Analyses
  and Alternatives to the Long-Range Ordered Water Hypothesis.
\newblock {\em The Journal of Physical Chemistry B} {\bf 2013}, {\em
  117},~7653--7674.

\bibitem[Pollack(2013)]{Pollack2013response}
Pollack, G.H.
\newblock Comment on {\textquotedblleft}A Theory of Macromolecular
  Chemotaxis{\textquotedblright} and {\textquotedblleft}Phenomena Associated
  with Gel{\textendash}Water Interfaces. Analyses and Alternatives to the
  Long-Range Ordered Water Hypothesis{\textquotedblright}.
\newblock {\em The Journal of Physical Chemistry B} {\bf 2013}, {\em
  117},~7843--7846.

\bibitem[Ovchinnikova and Pollack(2009)]{Ovchinnikova2009}
Ovchinnikova, K.; Pollack, G.H.
\newblock Can Water Store Charge?
\newblock {\em Langmuir} {\bf 2009}, {\em 25},~542--547.

\bibitem[Seger \em{et~al.}(2007)Seger, Vinodgopal, and Kamat]{Seger2007}
Seger, B.; Vinodgopal, K.; Kamat, P.V.
\newblock Proton Activity in Nafion Films:~ Probing Exchangeable Protons with
  Methylene Blue.
\newblock {\em Langmuir} {\bf 2007}, {\em 23},~5471--5476.

\bibitem[Phonyiem \em{et~al.}(2011)Phonyiem, Chaiwongwattana, Lao-ngam, and
  Sagarik]{Phonyiem2011}
Phonyiem, M.; Chaiwongwattana, S.; Lao-ngam, C.; Sagarik, K.
\newblock Proton transfer reactions and dynamics of sulfonic acid group in
  Nafion{\textregistered}.
\newblock {\em Physical Chemistry Chemical Physics} {\bf 2011}, {\em
  13},~10923.

\bibitem[Devanathan \em{et~al.}(2010)Devanathan, Venkatnathan, Rousseau,
  Dupuis, Frigato, Gu, and Helms]{Devanathan2010}
Devanathan, R.; Venkatnathan, A.; Rousseau, R.; Dupuis, M.; Frigato, T.; Gu,
  W.; Helms, V.
\newblock Atomistic Simulation of Water Percolation and Proton Hopping in
  Nafion Fuel Cell Membrane.
\newblock {\em The Journal of Physical Chemistry B} {\bf 2010}, {\em
  114},~13681--13690.

\bibitem[Kreuer \em{et~al.}(2000)Kreuer, Ise, Fuchs, and Maier]{Kreuer2000}
Kreuer, K.D.; Ise, M.; Fuchs, A.; Maier, J.
\newblock Proton and water transport in nano-separated polymer membranes.
\newblock {\em Le Journal de Physique {IV}} {\bf 2000}, {\em
  10},~Pr7--279--Pr7--281.

\bibitem[Musumeci and Pollack(2014)]{Musumeci2014}
Musumeci, F.; Pollack, G.H.
\newblock High electrical permittivity of ultrapure water at the
  water{\textendash}platinum interface.
\newblock {\em Chemical Physics Letters} {\bf 2014}, {\em 613},~19--23.

\bibitem[Faj{\'{\i}}n \em{et~al.}(2014)Faj{\'{\i}}n, Cordeiro, and
  Gomes]{Fajn2014}
Faj{\'{\i}}n, J.L.C.; Cordeiro, M.N.D.S.; Gomes, J.R.B.
\newblock Density Functional Theory Study of the Water Dissociation on Platinum
  Surfaces: General Trends.
\newblock {\em The Journal of Physical Chemistry A} {\bf 2014}, {\em
  118},~5832--5840.

\bibitem[Ninno(2017)]{DeNinno2017}
Ninno, A.D.
\newblock Dynamics of formation of the Exclusion Zone near hydrophilic
  surfaces.
\newblock {\em Chemical Physics Letters} {\bf 2017}, {\em 667},~322--326.

\bibitem[Hamaker(1937)]{Hamaker1937}
Hamaker, H.
\newblock The London{\textemdash}van der Waals attraction between spherical
  particles.
\newblock {\em Physica} {\bf 1937}, {\em 4},~1058--1072.

\bibitem[Lifshitz(1956)]{Lifshitz:1956}
Lifshitz, E.M.
\newblock {The theory of molecular attractive forces between solids}.
\newblock {\em Sov. Phys. JETP} {\bf 1956}, {\em 2},~73--83.

\bibitem[van Zwol and Palasantzas(2010)]{vanZwol2010}
van Zwol, P.J.; Palasantzas, G.
\newblock Repulsive Casimir forces between solid materials with
  high-refractive-index intervening liquids.
\newblock {\em Physical Review A} {\bf 2010}, {\em 81}.

\bibitem[Zhao \em{et~al.}(2011)Zhao, Koschny, Economou, and
  Soukoulis]{Zhao2011}
Zhao, R.; Koschny, T.; Economou, E.N.; Soukoulis, C.M.
\newblock Repulsive Casimir forces with finite-thickness slabs.
\newblock {\em Physical Review B} {\bf 2011}, {\em 83}.

\bibitem[Milling \em{et~al.}(1996)Milling, Mulvaney, and Larson]{Milling1996}
Milling, A.; Mulvaney, P.; Larson, I.
\newblock Direct Measurement of Repulsive van der Waals Interactions Using an
  Atomic Force Microscope.
\newblock {\em Journal of Colloid and Interface Science} {\bf 1996}, {\em
  180},~460--465.

\bibitem[Munday \em{et~al.}(2009)Munday, Capasso, and Parsegian]{Munday2009}
Munday, J.N.; Capasso, F.; Parsegian, V.A.
\newblock Measured long-range repulsive Casimir{\textendash}Lifshitz forces.
\newblock {\em Nature} {\bf 2009}, {\em 457},~170--173.

\bibitem[Meurk \em{et~al.}(1997)Meurk, Luckham, and Bergstr\"{o}m]{Meurk1997}
Meurk, A.; Luckham, P.F.; Bergstr\"{o}m, L.
\newblock Direct Measurement of Repulsive and Attractive van der Waals Forces
  between Inorganic Materials.
\newblock {\em Langmuir} {\bf 1997}, {\em 13},~3896--3899.

\bibitem[woo Lee and Sigmund(2002)]{Lee2002}
woo Lee, S.; Sigmund, W.M.
\newblock {AFM} study of repulsive van der Waals forces between Teflon
  {AF}{\texttrademark} thin film and silica or alumina.
\newblock {\em Colloids and Surfaces A: Physicochemical and Engineering
  Aspects} {\bf 2002}, {\em 204},~43--50.

\bibitem[Bevan and Prieve(1999)]{Bevan1999}
Bevan, M.A.; Prieve, D.C.
\newblock Direct Measurement of Retarded van der Waals Attraction.
\newblock {\em Langmuir} {\bf 1999}, {\em 15},~7925--7936.

\bibitem[woo Lee and Sigmund(2001)]{Lee2001}
woo Lee, S.; Sigmund, W.M.
\newblock Repulsive van der Waals Forces for Silica and Alumina.
\newblock {\em Journal of Colloid and Interface Science} {\bf 2001}, {\em
  243},~365--369.

\bibitem[Israelachvili(2011)]{Israelachvili2011}
Israelachvili, J.N.
\newblock {\em Intermolecular and Surface Forces}; Elsevier,  2011.

\bibitem[He \em{et~al.}(2018)He, Zhou, Wen, Shpilman, and Ren]{He2018}
He, X.; Zhou, Y.; Wen, X.; Shpilman, A.A.; Ren, Q.
\newblock Effect of Spin Polarization on the Exclusion Zone of Water.
\newblock {\em The Journal of Physical Chemistry B} {\bf 2018}, {\em
  122},~8493--8502.
\newblock
  doi:{\changeurlcolor{black}\href{https://doi.org/10.1021/acs.jpcb.8b04118}{\detokenize{10.1021/acs.jpcb.8b04118}}}.

\bibitem[Rodr{\'{\i}}guez-Fortu{\~{n}}o and Zayats(2016)]{RodrguezFortuo2016}
Rodr{\'{\i}}guez-Fortu{\~{n}}o, F.J.; Zayats, A.V.
\newblock Repulsion of polarised particles from anisotropic materials with a
  near-zero permittivity component.
\newblock {\em Light: Science {\&} Applications} {\bf 2016}, {\em
  5},~e16022--e16022.

\bibitem[Bunkin \em{et~al.}(2018)Bunkin, Shkirin, Kozlov, Ninham, Uspenskaya,
  and Gudkov]{Bunkin2018}
Bunkin, N.F.; Shkirin, A.V.; Kozlov, V.A.; Ninham, B.W.; Uspenskaya, E.V.;
  Gudkov, S.V.
\newblock Near-surface structure of Nafion in deuterated water.
\newblock {\em The Journal of Chemical Physics} {\bf 2018}, {\em 149},~164901.
\newblock
  doi:{\changeurlcolor{black}\href{https://doi.org/10.1063/1.5042065}{\detokenize{10.1063/1.5042065}}}.

\bibitem[Bunkin \em{et~al.}(2019)Bunkin, Bashkina, Bolikov, Bereza, Molchanov,
  and Kozlov]{Bunkin2019}
Bunkin, N.F.; Bashkina, U.A.; Bolikov, N.G.; Bereza, I.S.; Molchanov, I.I.;
  Kozlov, V.A.
\newblock Study of the luminescence from polymeric membrane swollen in water
  with various content of deuterium; isotopic effects.
\newblock {\em Journal of Physics: Conference Series} {\bf 2019}, {\em
  1348},~012030.
\newblock
  doi:{\changeurlcolor{black}\href{https://doi.org/10.1088/1742-6596/1348/1/012030}{\detokenize{10.1088/1742-6596/1348/1/012030}}}.

\bibitem[Almeida and Kawano(1997)]{DeAlmeida1997}
Almeida, S.H.D.; Kawano, Y.
\newblock Ultraviolet-visible spectra of Nafion membrane.
\newblock {\em European Polymer Journal} {\bf 1997}, {\em 33},~1307--1311.
\newblock
  doi:{\changeurlcolor{black}\href{https://doi.org/10.1016/s0014-3057(96)00217-0}{\detokenize{10.1016/s0014-3057(96)00217-0}}}.

\bibitem[Horinek \em{et~al.}(2008)Horinek, Serr, Geisler, Pirzer, Slotta, Lud,
  Garrido, Scheibel, Hugel, and Netz]{Horinek2008}
Horinek, D.; Serr, A.; Geisler, M.; Pirzer, T.; Slotta, U.; Lud, S.Q.; Garrido,
  J.A.; Scheibel, T.; Hugel, T.; Netz, R.R.
\newblock Peptide adsorption on a hydrophobic surface results from an interplay
  of solvation, surface, and intrapeptide forces.
\newblock {\em Proceedings of the National Academy of Sciences} {\bf 2008},
  {\em 105},~2842--2847.

\bibitem[Richert \em{et~al.}(2011)Richert, Agapov, and Sokolov]{Richert2011}
Richert, R.; Agapov, A.; Sokolov, A.P.
\newblock Appearance of a Debye process at the conductivity relaxation
  frequency of a viscous liquid.
\newblock {\em The Journal of Chemical Physics} {\bf 2011}, {\em 134},~104508.

\bibitem[{Chaplin}(2012)]{2012arXivChaplin}
{Chaplin}, M.
\newblock {Self-generation of colligative properties at hydrophilic surfaces}.
\newblock {\em arXiv e-prints: 1203.0206} {\bf 2012}.

\bibitem[Chaplin()]{chaplinweb}
Chaplin, M.
\newblock Self-generation of colligative properties at interfaces.
\newblock \url{http://www1.lsbu.ac.uk/water/colligative_generation.html}.
\newblock Accessed: 2019-09-14.

\bibitem[Rohani and Pollack(2013)]{Rohani2013}
Rohani, M.; Pollack, G.H.
\newblock Flow through Horizontal Tubes Submerged in Water in the Absence of a
  Pressure Gradient: Mechanistic Considerations.
\newblock {\em Langmuir} {\bf 2013}, {\em 29},~6556--6561.
\newblock
  doi:{\changeurlcolor{black}\href{https://doi.org/10.1021/la4001945}{\detokenize{10.1021/la4001945}}}.

\end{thebibliography}



\end{document}